\documentclass[12pt]{article}
\usepackage{amsmath}
\usepackage{graphicx}
\usepackage[hyphens]{url}
\usepackage{hyperref}
\usepackage{natbib}
\usepackage{booktabs}
\usepackage{geometry}
\newcommand{\wref}[1]{\textsuperscript{[W#1]}}

\title{Binding Constraints in a Strategic Technology:\\ An Industrial Vulnerability Index for the Foundation Model Industry}

\author{%
Claudio Pirrone\thanks{Corresponding author. Department of Business, Economics, and Statistics, University of Palermo. Email: \texttt{claudio.pirrone@unipa.it}.}
\and Stefano Fricano\thanks{Department of Business, Economics, and Statistics, University of Palermo. Email: \texttt{stefano.fricano@unipa.it}.}
\and Gioacchino Fazio\thanks{Department of Business, Economics, and Statistics, University of Palermo. Email: \texttt{gioacchino.fazio@unipa.it}.}
}
\date{}

\begin{document}
\maketitle

\section*{Highlights} \begin{itemize} \item We develop the Artificial Intelligence Industrial Vulnerability Index (AIIVI) to assess input constraints in the foundation model industry. \item Using an O-Ring aggregation framework, the index identifies the inputs whose scarcity most strongly limits industry capability. \item Empirical estimates for frontier developers show that compute and energy are the dominant sources of vulnerability and are largely interchangeable at the margin. \item The analysis reveals three underappreciated risks: external-debt financing of data centres, local opposition to infrastructure siting, and model-layer export controls. \end{itemize}

\begin{abstract}
Industrial policy directed at artificial intelligence has become substantial. Public investment concentrates on compute, where private capacity expansion is already underway, and is largely absent from grid interconnection and generation. Whether that allocation accurately reflects the most binding constraint in production has not yet been systematically measured. This paper introduces the Artificial Intelligence Industrial Vulnerability Index (AIIVI) as a framework for assessing potential production constraints in frontier AI. We construct a composite index of input vulnerability for the foundation model industry from fourteen concentration and sustainability parameters, aggregated by an O-Ring functional form in which capability is strongly influenced by the most constrained input. 
AIIVI should not be interpreted as a direct measurement of production constraints. Rather, it provides an index-based approximation derived from observable indicators and weighting assumptions designed to capture relative exposure to input vulnerabilities.
Using this transparent and reproducible framework for estimating the relative vulnerability of frontier AI production to disruptions across key input dimensions, we also evaluate the index at 31 August 2026 for the nine developers at the capability frontier. We find values of 0.75 under AI-elicited weights and 0.62 under equal weights. The analysis identifies compute and energy as the principal drivers of input vulnerability. Marginal changes in these dimensions produce substantially larger variations in AIIVI than comparable changes in capital, talent, or data. We highlight three exposures that receive limited attention in existing assessment frameworks: the increasing reliance on external debt for data-centre financing, local opposition as a constraint on data-centre siting, and export controls targeting deployed AI models rather than hardware.

\end{abstract}

\noindent\textbf{Keywords:} Artificial Intelligence Industry, vulnerability, composite indicators, foundation models\\[4pt]
\noindent\textbf{JEL codes:} O25, O33, L52, L86, F52

\section{Introduction}

Industrial policy directed at artificial intelligence has become substantial. The United States has committed \$52.7 billion to semiconductor manufacturing, research and workforce development under the CHIPS Act, the European Union 43 billion euros under its Chips Act, and export controls on advanced semiconductors are now a leading instrument of technological competition \citep{Miller2022, Farrell2019}.\wref{1} These interventions share a diagnostic premise: that the input they strengthen is the one constraining production.

Frontier model production has a definite physical structure. A training run occupies a cluster of tens of thousands of accelerators, fabricated at leading-edge nodes and packaged with high-bandwidth memory, for a period of weeks. The cluster draws power at the scale of a small city and must be sited where that power can be delivered. It is financed years before it earns revenue. It is designed and supervised by a research staff numbered in the hundreds worldwide. And it consumes a corpus whose scale and quality bound the capability obtained.

Advanced semiconductors, energy infrastructure, capital, specialised talent and training data are therefore not a list of inputs but a sequence of requirements, each of which must be met for the others to matter. Available evidence suggests limited substitutability among these inputs within policy-relevant time horizons: subsidised fabrication without electricity to run the resulting chips adds no capability. Under such complementarity, capability is strongly influenced by the most constrained input \citep{Kremer1993}, and the return to intervention depends on whether it relaxes the constraint that binds \citep{HausmannRodrikVelasco2008}. Which input constitutes the binding constraint remains largely unexplored at the industry level.

Measurement is obstructed by the object. Leading developers are private and disclose little; market structure changes within quarters; official statistics do not report at this frequency or coverage. We construct the Artificial Intelligence Industrial Vulnerability Index (AIIVI) from fourteen concentration and sustainability parameters, each traceable to a primary or named source, aggregated by an O-Ring functional form. We estimate the index at 31 August 2026 for the nine developers at the capability frontier, under two weight specifications: equal weights, and weights elicited at the reference date from an AI system. All sources are publicly accessible, permitting re-estimation at the industry's own frequency.

We find an index of 0.75 under elicited weights and 0.62 under equal weights. Compute (0.83) and energy (0.86) dominate the remaining dimensions and are close to interchangeable at the margin: a ten percent reduction in either lowers the index by about five percent, against 0.4 percent for capital, 0.3 percent for talent and 0.1 percent for data (Table~\ref{tab:marginal}). The ranking holds under both specifications, and it does not correspond to the allocation of policy effort.

Three developments within the reference period are not priced by existing frameworks: the migration of data-centre financing from internally generated cash to external debt, local opposition as a binding constraint on siting, and export controls applied to a deployed model rather than to hardware. A further implication may be relevant for European policy, since no European developer reaches the frontier set, the leading one scoring twenty-nine points below the frontier average.

Section 2 reviews the literature. Section 3 constructs the index. Section 4 describes data and measurement. Section 5 reports results. Section 6 draws out policy implications. Section 7 concludes.

\section{Literature review}

Three strands of literature motivate our approach.

The first concerns complementarity in production. Systems composed of non-substitutable activities are governed by their bottlenecks rather than by average input quality \citep{Hirschman1958}, a logic formalised in the O-Ring production function, where output is multiplicative in task quality \citep{Kremer1993}. The macroeconomic literature on production networks establishes the same property at the level of an economy: sectoral shocks need not wash out in aggregate when linkages are concentrated \citep{Acemoglu2012, Carvalho2019}, and where inputs are complements the aggregate response is not given by the affected sector's sales share, since Hulten's theorem fails at second order with losses concentrated where substitution is hardest \citep{Baqaee2019}. Applications have addressed labour, firm organisation and whole economies. We extend the reasoning to an industry characterised by limited substitutability among key inputs.

The second concerns the diagnosis of constraints. \citet{HausmannRodrikVelasco2008} turn bottleneck logic into a policy method, ranking reforms by the shadow price of the constraint they relax. The empirical literature on industrial policy reaches a compatible conclusion from the other direction, finding effects conditional on the fit between instrument and structural context \citep{Rodrik2004, Aghion2015, JuhaszLaneRodrik2024}. Establishing that fit requires knowing which constraint binds. Growth diagnostics is applied at country level, to aggregate growth; to the best of our knowledge it has no counterpart at the level of a single industry, although the industry is the unit at which contemporary technology policy operates.

The third concerns strategic dependence in production networks. \citet{Farrell2019} show how network chokepoints become instruments of statecraft, \citet{Miller2022} traces this in semiconductors, and \citet{Gereffi2005} and \citet{Baldwin2022} study governance and risk in value chains more generally. These studies identify where exposure lies without ranking exposures against one another.

The closest methodological relatives lie outside these strands, in criticality assessments of material inputs. These score each material on two dimensions, the likelihood of a supply disruption and its consequence for the economy that depends on the material \citep{NRC2008, Graedel2012, Blengini2017}, and they underpin the European Union's list of critical raw materials. Our index shares that structure: parameters measure exposure, weights measure consequence. It differs in the object, an industry's inputs rather than an economy's materials, and in the aggregation, since criticality assessments place materials in a two-dimensional matrix and read criticality off the position, while we combine the dimensions multiplicatively as complementarity requires. Composite indicators supply the construction methodology, covering normalisation, weighting and robustness \citep{Nardo2008}. Weighting is the contested step: most vulnerability indices impose equal weights or derive them from expert panels, which are costly to reconvene. We report two specifications, one equal-weighted and one elicited at the reference date from an AI system, and test the sensitivity of results to both.

Our contribution connects these literatures. We use the logic of bottlenecks, the diagnostic structure of growth diagnostics and the empirical concern with strategic dependence to construct an industry-level vulnerability index. In doing so we shift attention from the diffusion of foundation models as a general purpose technology \citep{BresnahanTrajtenberg1995, Agrawal2019} to the concentrated production system that makes them possible.

\section{The index}

Let frontier capability $Y$ depend on five inputs indexed $i \in \{C,E,T,K,D\}$ for compute, energy, talent, capital and data. Write $a_i \in [0,1]$ for the \emph{availability} of input $i$, the share of the required supply that can be obtained. Following the O-Ring structure, inputs enter multiplicatively:
\begin{equation}
Y = B \prod_i a_i^{\alpha_i}, \qquad \sum_i \alpha_i = 1,
\end{equation}
where $B$ collects technology and organisation and $\alpha_i$ is the elasticity of capability with respect to the availability of input $i$. Constant returns in availabilities normalise $Y = B$ when every input is fully available.

Define the vulnerability of input $i$ as the share of its supply exposed to impairment,
\begin{equation}
V_i = 1 - a_i \in [0,1].
\end{equation}
Substituting and dividing by capability under full availability gives $Y/B = \prod_i (1-V_i)^{\alpha_i}$, and the index is its complement:
\begin{equation}
\text{AIIVI} = 1 - \frac{Y}{B} = 1 - \prod_i (1-V_i)^{\alpha_i}.
\end{equation}
AIIVI may be interpreted as an indicator of the productive capability potentially exposed to input vulnerabilities. A value of 0.75 indicates that a substantial share of frontier production capability depends on inputs exhibiting significant vulnerability.

Three properties follow from (6). Any single input approaching $V_i = 1$ drives the index toward unity irrespective of the others, which is the weakest-link result. Inputs are complements, $\partial^2 Y/\partial a_i \partial a_j > 0$, so the marginal value of improving one input rises with the availability of the rest. And the marginal effect of dimension $i$ on the index,
\begin{equation}
\frac{\partial \text{AIIVI}}{\partial V_i} = \frac{\alpha_i\,(1-\text{AIIVI})}{1-V_i},
\end{equation}
rises in both $\alpha_i$ and $V_i$: criticality and observed exposure jointly determine where relaxation pays. Section 5.2 evaluates (7) at the estimated values.

Equation (4) also fixes the meaning of $\alpha_i$ for elicitation. Reducing the availability of input $i$ by a proportion $\lambda$ multiplies capability by $\lambda^{\alpha_i}$: at $\alpha_C = 0.35$, halving compute availability leaves 78 percent of capability. Respondents judge this quantity directly, so the weights are elicited on the scale at which they enter the model.

The weights price the consequence of an input reduction, not its likelihood, which is carried by the parameters. The index is accordingly a structural diagnostic rather than a forecast.

Each dimension aggregates observable parameters linearly:
\begin{equation}
V_i = \sum_{j \in J_i} w_{ij}\,x_{ij}, \qquad \sum_{j \in J_i} w_{ij} = 1.
\end{equation}
Concentration parameters take $\sqrt{\sum_k s_k^2}$ over market shares $s_k$, preserving comparability across markets of differing granularity. Ratio parameters are capped at one. Parameters computed across firms use industry totals rather than averages of firm-level ratios, so that firms enter in proportion to economic weight. Growth ratios are normalise to one when demand growth exceeds supply growth, reflecting an assumption of maximum exposure.

\begin{table}[htbp]
\centering\small
\caption{Dimensions and parameters}\label{tab:channels}
\begin{tabular}{llp{7.2cm}}
\toprule
\textbf{Dimension} & \textbf{Parameters} & \textbf{Exposure measured} \\
\midrule
Compute & $SC$, $GEO$, $PR$ & Fabrication, geographic and design concentration \\
Energy  & $GR_{e}$, $GR_{c}$, $EB$ & Demand--supply growth imbalance; efficiency deceleration \\
Talent  & $E$, $RC$ & Elite concentration; corporate brokerage of knowledge flows \\
Capital & $DEP$, $SU$, $RI$, $VS$ & Investor lock-in; burn sustainability; valuation risk; funding concentration \\
Data    & $R$, $C$ & Quality headroom; training-cost economics \\
\bottomrule
\end{tabular}
\end{table}

\section{Data and measurement}

\subsection{Data and sample}

All estimates refer to 31 August 2026. Economic parameters use the most recent reporting period ended on or before that date, taking the disclosure as published. Model scores are those of the leaderboard provider for models released by that date \citep{ArtificialAnalysis2026}.

We define the frontier as models within ten index points of the leader. The convention in the model-tracking literature is a top-$N$ cut on training compute at release \citep{EpochAI2026}; training compute is not disclosed for the models in our sample, so we apply the analogous cut on measured capability. Nine models from nine developers qualify, one per developer (Table~\ref{tab:models}). Their average is 58.8, giving $R = 0.412$. An eight-point gap separates the ninth model from the tenth, so thresholds between ten and seventeen points select the same set.

Three features of the sample bear on interpretation. Meta re-entered the frontier during 2026 after rebuilding its research organisation, and its frontier model is proprietary, departing from its earlier open-weights strategy. Moonshot and DeepSeek publish open weights at the frontier. No European developer reaches the frontier set; the leading one scores thirty. Firms outside the set are excluded throughout.

Parameters are assembled from company filings and releases, institutional statistics, and named outlets, through an AI-assisted pipeline in which the model extracts, computes and drafts, and the authors validate or reject each output against the original source. Large language model sessions are not deterministic \citep{Ouyang2025}, so reproducibility rests on the data, source catalogue and code rather than on the prompts; these are provided in Supplementary Materials 1--3.

Sources outside the peer-reviewed and official-statistical literature are used throughout, since neither reports at the frequency or coverage the object requires. Claims resting on such sources are footnoted with publisher, exact URL and date of retrieval; where a secondary source reports a figure originating with a named institution, both are given. Tier gradings enter the confidence rating of each parameter.

\subsection{Parameters}

Table~\ref{tab:params} reports the fourteen parameters and their confidence ratings. Sources and derivations for all of them are catalogued in Supplementary Material 2. Four constructions require comment.

Design concentration is computed from disclosed data-centre segment revenue at NVIDIA, Broadcom, AMD and Marvell, whose reporting periods end within a five-week span. The measure spans the merchant vendor and the suppliers of custom accelerators, which is where substitution operates: four sample developers have custom silicon in production, designed with those suppliers. Revenue is used as proxy for deployed compute, acknowledging the limitations associated with indirect measurement.

Strategic dependency covers unconditional strategic equity from any source rather than cloud providers alone. Operating sustainability and valuation risk are computed on industry totals. Energy-efficiency deceleration is measured along the general-purpose accelerator line, which governs frontier training; specialised inference architectures lie outside it.

\begin{table}[htbp]
\centering\small
\caption{Parameter values at 31 August 2026}\label{tab:params}
\begin{tabular}{llcc}
\toprule
\textbf{Dimension} & \textbf{Parameter} & \textbf{Value} & \textbf{Confidence} \\
\midrule
Compute & $SC$ & 0.732 & High \\
        & $GEO$ & 0.920 & High \\
        & $PR$ & 0.793 & Medium \\
Energy  & $GR_{e}$ & 1.000 & High \\
        & $GR_{c}$ & 1.000 & High \\
        & $EB$ & 0.533 & Medium \\
Talent  & $E$ & 0.458 & Medium-Low \\
        & $RC$ & 0.479 & High \\
Capital & $DEP$ & 0.590 & Medium \\
        & $SU$ & 0.104 & Low \\
        & $RI$ & 0.316 & Medium-High \\
        & $VS$ & 0.465 & High \\
Data    & $R$ & 0.412 & High \\
        & $C$ & 0.049 & Medium \\
\bottomrule
\end{tabular}
\end{table}

\subsection{Weights}

Scenario A uses weights elicited from Claude Opus 5 as an exploratory benchmark for assessing perceived input criticality, selected as the highest-scoring model accessible to us on the economics-specific capability index of the leaderboard provider, where it scores 63 of 100. The rule is stated so that a later estimation can apply it to whichever model then leads. Cadence motivates the design: an index re-estimated at the industry's frequency requires a weighting method of matching frequency. An approach of this kind is supported by evidence on language models as expert respondents and forecasters \citep{Horton2023, Argyle2023, Aher2023, Schoenegger2024}. Weights are integer multiples of 0.05. The instrument, the responses, and the eliciting model's conflict of interest --- it is a product of a sample firm --- are in Supplementary Material 3. Scenario B applies equal weights within and across dimensions.

The instrument was piloted with human experts and frontier AI systems answering identically \citep{Dalkey1963, Linstone1975}. Their weight sets differ systematically, the AI respondents tended to assign slightly higher vulnerability scores, averaging 0.07 points higher on the index scale, but all place the index between 0.59 and 0.82. Details are available from the authors on request.

\begin{table}[htbp]
\centering\small
\caption{Scenario A weights (elicited, 31 August 2026)}\label{tab:wA}
\begin{tabular}{llp{7.4cm}}
\toprule
 & \textbf{Weight} & \textbf{Rationale} \\
\midrule
\multicolumn{3}{l}{\textit{Compute}} \\
$GEO$ & 0.50 & No geographic substitute for leading-edge capacity \\
$SC$  & 0.35 & Custom accelerators substitute away from the designer, not the fab \\
$PR$  & 0.15 & Four sample developers have custom silicon in production \\
\multicolumn{3}{l}{\textit{Energy}} \\
$GR_{e}$ & 0.60 & Grid access carries a permitting component with no substitute \\
$GR_{c}$ & 0.10 & Carbon constraint displaced by energy-security priority \\
$EB$     & 0.30 & Efficiency trajectory is the swing variable \\
\multicolumn{3}{l}{\textit{Talent}} \\
$E$  & 0.50 & Evidence suggest that acquisition activity may have contributed to the rapid rebuilding of frontier capabilities. \\
$RC$ & 0.50 & Two sample developers publish open weights \\
\multicolumn{3}{l}{\textit{Capital}} \\
$DEP$ & 0.40 & Lock-in through supplier equity and credit support \\
$SU$  & 0.25 & Cash-flow fundamental \\
$VS$  & 0.20 & Debt has displaced venture at the funding margin \\
$RI$  & 0.15 & Multiples adjustable; listing opens a correction channel \\
\multicolumn{3}{l}{\textit{Data}} \\
$R$ & 0.70 & Quality headroom \\
$C$ & 0.30 & Cost economics at current capital intensity \\
\multicolumn{3}{l}{\textit{Criticality}} \\
\multicolumn{3}{l}{Compute 0.35; Energy 0.30; Capital 0.20; Talent 0.10; Data 0.05} \\
\bottomrule
\end{tabular}
\end{table}

\section{Results}

\subsection{Estimates}

Table~\ref{tab:results} reports the index and its components. AIIVI is 0.754 under elicited weights and 0.625 under equal weights, with Monte Carlo intervals of $[0.699, 0.814]$ and $[0.581, 0.677]$. Energy (0.86) and compute (0.83) exceed the remaining dimensions under both specifications; talent, capital and data fall between 0.23 and 0.47.

\begin{table}[htbp]
\centering\small
\caption{AIIVI at 31 August 2026}\label{tab:results}
\begin{tabular}{lcc}
\toprule
 & \textbf{Scenario A} & \textbf{Scenario B} \\
 & \textit{elicited weights} & \textit{equal weights} \\
\midrule
$V_E$ (Energy)  & 0.860 & 0.844 \\
$V_C$ (Compute) & 0.834 & 0.814 \\
$V_T$ (Talent)  & 0.469 & 0.469 \\
$V_K$ (Capital) & 0.402 & 0.369 \\
$V_D$ (Data)    & 0.303 & 0.230 \\
\midrule
\textbf{AIIVI} & \textbf{0.754} & \textbf{0.625} \\
Monte Carlo 95\% CI & [0.699, 0.814] & [0.581, 0.677] \\
\bottomrule
\end{tabular}
\end{table}

Equation (2) gives the marginal effect of each dimension at the estimated values (Table~\ref{tab:marginal}).

\begin{table}[htbp]
\centering\small
\caption{Marginal effects by dimension, Scenario A}\label{tab:marginal}
\begin{tabular}{lcccc}
\toprule
 & $V_i$ & $\partial A/\partial V_i$ & Elasticity & $A$ at $V_i=0$ \\
\midrule
Compute & 0.834 & 0.519 & 0.574 & 0.539 \\
Energy  & 0.860 & 0.527 & 0.601 & 0.556 \\
Capital & 0.402 & 0.082 & 0.044 & 0.727 \\
Talent  & 0.469 & 0.046 & 0.029 & 0.738 \\
Data    & 0.303 & 0.018 & 0.007 & 0.750 \\
\bottomrule
\end{tabular}
\end{table}

Compute and energy emerge as the most influential dimensions and display similar marginal effects on the index. A ten percent reduction in either lowers the index by about five percent. The same reduction applied to capital, talent or data moves it by less than half a percent.

The two differ in total rather than marginal terms. Eliminating compute vulnerability entirely gives an index of 0.54, eliminating energy vulnerability 0.56: compute carries the higher criticality weight, energy the higher observed vulnerability. Neither elimination brings the index below the high range, which follows from complementarity, since relaxing one constraint leaves the others binding.

The asymmetry lies in the allocation of effort. Compute attracts fabrication subsidies on three continents and a private investment cycle at record scale. Energy attracts no instrument of comparable size, and its two growth parameters sit at their normalised maximum.

\subsection{The dimensions}

\emph{Compute.} Concentration is high on three channels: 72.5 percent of the foundry market in one firm \citep{TrendForce2026}, 92 percent of leading-edge capacity in one location, and 78 percent of disclosed data-centre silicon revenue in one vendor \citep{NVIDIA2026, AMD2026}. Entry is visible on the design channel. Custom accelerator programmes reached production at Google, Meta, OpenAI and Anthropic; AMD grew data-centre revenue 57 percent year on year \citep{AMD2026}; NVIDIA licensed Groq's inference technology and engineering team for \$20 billion in December 2025.\wref{2} Those custom parts are fabricated by the same foundry, so design substitution leaves the fabrication channel unaffected.

\emph{Energy.} Both growth ratios stand at their cap: electricity demand from AI-focused data centres grew by half in 2025 against three percent for the grid, and the International Energy Agency estimates that a fifth of planned projects face grid-connection delays \citep{IEA2026kq, IEA2026elec}. Capital expenditure by the four largest infrastructure providers was \$413 billion in 2025 and is projected at approximately \$760 billion for 2026.\wref{3} The efficiency parameter measures the general-purpose accelerator line; specialised inference architectures, where reported cost per token is a fraction of general-purpose parts, lie outside it.

\emph{Talent.} Five corporate laboratories broker 0.48 of the structural knowledge flow in frontier research, one of them accounting for half of that. The two largest commercial developers are near-absent from the co-authorship network, holding their advantage as tacit knowledge. Meta appears to have regained a frontier position within twelve months after substantial organisational and talent investments, a \$14.3 billion transaction for Scale AI bringing in its chief executive and senior staff \citep{Meta2026}, which bears on how far capital substitutes for accumulated expertise.

\emph{Capital.} Strategic dependency is 0.59 over unconditional strategic equity: NVIDIA holds 45 percent, Amazon 32 percent and Microsoft 21 percent \citep{OpenAI2026round, Anthropic2026, Bloomberg2026}. NVIDIA also reported \$279 billion of supply and capacity commitments at 26 July 2026 and, in August 2026, entered guarantees capped at \$105 billion providing credit support for data-centre leases hosting OpenAI workloads.\wref{4} Those guarantees are phased from fiscal 2029 as facilities reach service-ready status and terminate if OpenAI attains an investment-grade rating; being contingent, they fall outside the parameter. Developers absorbed \$237 billion of the \$510 billion in global venture funding in the first half of 2026 \citep{Crunchbase2026}. Aggregate capitalisation is 25 times aggregate annualised revenue, and aggregate burn a tenth of four years' revenue, on divergent trajectories: OpenAI projects burning 57 percent of revenue through 2027, Anthropic roughly a third in 2026 falling to single digits \citep{DeveloperProjections2026}.

\emph{Data.} The frontier averages 59 of 100 on economically relevant tasks. Training-cost economics remain manageable relative to revenue at the leaders; the parameter covers training rather than inference cost.

\subsection{Robustness}

Every variant lies between 0.68 and 0.79 (Table~\ref{tab:robust}). Pilot-panel weights give 0.681. Excluding the two parameters rated below medium confidence raises the index by 0.009. The frontier cutoff is immaterial across all thresholds examined. Computing the capital ratios as firm averages rather than industry totals moves the index by 0.001. Two choices matter more: the efficiency parameter's alternative bases span 0.726 to 0.770, and the design-concentration parameter differs by 0.011 between the four-supplier and two-vendor constructions.

\begin{table}[htbp]
\centering\small
\caption{Robustness, Scenario A unless noted}\label{tab:robust}
\begin{tabular}{lc}
\toprule
Pilot-panel weights ($n{=}11$; $n{=}13$) & 0.681; 0.699 \\
Medium-or-above confidence parameters only & 0.763 \\
$EB$ alternative bases (0.333 / 0.627) & 0.726 / 0.770 \\
$PR$ on the merchant market alone (0.931) & 0.765 \\
$DEP$ restricted to cloud providers (0.693) & 0.758 \\
$SU$ and $RI$ as firm averages & 0.755 \\
$R$: five-point cutoff (0.395) / eleven-model set (0.451) & 0.754 / 0.755 \\
\bottomrule
\end{tabular}
\end{table}

\section{Policy implications}

\subsection{Allocation across constraints}

Under complementarity, the return to public effort depends on the vulnerability of the input it addresses and on how much of that vulnerability private action is already relaxing. Our estimates place compute and energy close to interchangeable at the margin, and both an order of magnitude above the remaining dimensions. They differ in the effort already directed at them. Semiconductor fabrication receives the public commitments cited in Section 1, alongside a private investment cycle at record scale. The energy dimension sits at its normalised maximum on both growth parameters, and no instrument of comparable size addresses interconnection queues, permitting, transmission or firm generation. The results suggest that additional investments in energy infrastructure may generate marginal benefits comparable to those currently sought through semiconductor support policies.

\subsection{Unpriced exposures}

Gallup, surveying 1{,}000 US adults between 2 and 18 March 2026, found 71 percent opposed to data-centre construction in their local area, 48 percent strongly opposed, against 53 percent opposing a local nuclear plant.\wref{5} Monitoring Analytics, the independent market monitor for the PJM Interconnection, attributed 63 percent of the increase in the 2025/2026 capacity auction to data-centre load, some \$9.3 billion recovered from customers across thirteen states in one year; capacity prices moved from \$28.92 per megawatt-day in 2024/2025 to \$329.17 in 2026/2027.\wref{6}

Restrictions followed within the reference period. On 14 July 2026 the Governor of New York signed Executive Order 62, the first statewide moratorium on new hyperscale data centres, pausing discretionary environmental permits for facilities of 50 megawatts or more for up to a year.\wref{7} In Texas, the Governor directed ERCOT and state regulators to audit an interconnection queue of roughly 474 gigawatts, some 90 percent of it data centres, and stated on 20 August 2026 that the directive had halted up to 1{,}800 projects.\wref{8}

The mechanism runs from buildout to retail prices to political constraint. We do not include it in the index: the evidence is almost entirely from one jurisdiction, and siting constraints elsewhere operate through different channels, state direction in China and grid-connection policy in Europe. It belongs in the policy account because it converts electricity from a cost input into a question of social licence.

Data-centre investment has moved onto balance sheets that formerly funded it internally. Alphabet reported negative free cash flow of \$5.9 billion for the second quarter of 2026, its first negative quarter since listing in 2004, on capital expenditure of \$44.9 billion.\wref{9} Meta reported operating cash flow of \$31.86 billion against capital expenditure of \$31.08 billion, leaving free cash flow of \$784 million.\wref{10} Across the five largest providers, FactSet estimates that incremental annual debt rose from 9 percent of capital spending in fiscal 2024 to 32 percent on a trailing basis by mid-2026, and dollar debt outstanding more than doubled to over \$360 billion.\wref{11}

Credit markets are pricing that shift. Of ninety-one hyperscaler bonds issued in 2026 with comparable pricing data, seventy-eight traded at higher yields on 28 July than at issue, a median increase of about 22 basis points. In July 2026, investors sought yields above 7 percent on a \$12 billion financing for Meta's El Paso campus, raised through a joint venture with BlackRock, roughly 0.4 percentage points above the terms obtained on the comparable \$27 billion Hyperion financing nine months earlier.\wref{12} Lenders are differentiating between facilities rather than pricing them off the sponsor's name.

The index carries no financing-conditions parameter, since continuous measures of availability and price are not obtainable from primary sources. What it measures is the dependency on which credit conditions act. Competition authorities and financial-stability bodies have reason to monitor this channel at the frequency they apply to the input constraints.

Export controls have operated on inputs: lithography, accelerators, fabrication capacity. In June 2026 they were applied to a deployed frontier model, suspended worldwide for eighteen days and restored on withdrawal of the directive. Control at the model layer acts on knowledge rather than hardware, applies by nationality rather than geography, and can bind firms in the controlling jurisdiction. The reclassification is symmetric: a leading Chinese developer has been on the US Entity List since January 2025, a leading US developer operates under a defence supply-chain designation, and frontier laboratories publish dual-use safety evaluations as a component of release. We keep dual-use status outside the index, as a property of the governance environment rather than of input substitutability. Its policy implication belongs here: an instrument that revokes access to a production input on a timescale of days should be assessed within the same complementarity framework as subsidy.

\subsection{The European position}

No European developer reaches the frontier set. The leading one scores thirty against a frontier average of 58.8, and would not qualify under any cutoff we examine. The financing gap runs alongside it: OpenAI and Anthropic raised \$122 billion and \$65 billion in single rounds during the first half of 2026 \citep{OpenAI2026round, Anthropic2026}, each exceeding by an order of magnitude the cumulative external funding raised by any European developer. Complementarity implies that partial strategies buy little: capability requires the joint availability of compute, energy, capital, talent and data, and the European deficit is simultaneous across at least three. The findings suggest potential benefits from a more integrated approach linking grid interconnection, compute procurement and capital-market development.

Every parameter in the index is constructible from sources that are free to access: quarterly corporate disclosures, institutional statistical releases, research-firm press summaries and a public capability leaderboard. No parameter depends on a subscription product. The marginal cost of a full estimation is measured in tens of person-hours, which makes quarterly re-estimation feasible for a policy apparatus committing tens of billions annually.

What a series would track is structural rather than incremental. Within the observation window alone, custom accelerator programmes reached production at four of the nine sample developers, a chip supplier became the largest holder of unconditional strategic equity in the frontier, and the financing of data-centre construction moved from internally generated cash to external debt. Each bears on a parameter of the index, and none is visible in annual statistics.

\section{Conclusions}

Contemporary industrial policy toward artificial intelligence rests on a premise about which input constrains production. This paper proposes a framework for measuring the relative importance of such constraints. We construct a fourteen-parameter composite index of input vulnerability for the foundation model industry, aggregated by an O-Ring functional form, and estimate it at 31 August 2026 for the nine developers at the capability frontier.

The index is 0.75 under elicited weights and 0.62 under equal weights, with every variant examined between 0.68 and 0.79. Compute and energy dominate the remaining dimensions and are close to interchangeable at the margin: a ten percent reduction in either lowers the index by about five percent, against less than half a percent for capital, talent and data. They differ in the effort directed at them, with public and private investment concentrated on compute and largely absent from energy.

Our contribution is the diagnostic layer. Growth diagnostics rank reforms by the shadow price of the constraint they relax; we apply that logic at the level of an industry, at a frequency that matches the industry's rate of change. The reference period also produced three exposures that existing frameworks do not price: the migration of data-centre financing onto external debt, the emergence of local opposition as a constraint on siting, and the application of export controls to a deployed model.

Three limitations bear on interpretation. Two parameters are rated below medium confidence, and excluding both moves the index by 0.009. Design concentration is measured on disclosed revenue rather than deployed capacity, which no vendor discloses, and it carries the largest definitional effect in the paper. The weights come from a single AI elicitor with a disclosed conflict of interest; the instrument is published so that cross-vendor re-elicitation can corroborate or overturn them. The index measures structure at a point in time, and its weights price the impact of an input reduction rather than its likelihood. Events acting on that structure fall outside the measure. Two extensions would widen it: a fuel-supply-security component for the energy dimension, and a measure of exposure to control at the model layer.

\section*{Declarations}

\noindent\textbf{Funding.} This work received no external funding.

\medskip\noindent\textbf{Declaration of competing interest.} The authors declare that they have no known competing financial interests or personal relationships that could have appeared to influence the work reported in this paper. The study employs commercial AI systems both as research instruments and as objects of study; the authors have no financial relationship with any of these firms.

\medskip\noindent\textbf{Data availability.} The raw elicitation responses, the co-authorship network extraction scripts and outputs, the source catalogues with verbatim references, and replication code are provided as supplementary materials. 

\medskip\noindent\textbf{Declaration of generative AI and AI-assisted technologies.} Generative AI is used in two capacities. As a research instrument, AI-assisted extraction and AI-based weight elicitation are part of the methodology, described in Sections 3 and 4, with every extracted value validated by the authors against primary sources and all instruments disclosed in Supplementary Material 3. As writing assistance, the authors used Claude (Anthropic) to assist with drafting, editing, computational verification and \LaTeX{} preparation. The authors reviewed and edited the content and take full responsibility for the published article.

\medskip\noindent\textbf{CRediT authorship contribution statement.} Authors equally contributed to this work.

\bibliographystyle{elsarticle-harv}
\bibliography{bib2}

@article{Kremer1993,
  author    = {Kremer, Michael},
  title     = {The O-Ring Theory of Economic Development},
  journal   = {The Quarterly Journal of Economics},
  volume    = {108},
  number    = {3},
  pages     = {551--575},
  year      = {1993},
  doi       = {10.2307/2118400}
}

@book{Miller2022,
  author    = {Miller, Chris},
  title     = {Chip War: The Fight for the World's Most Critical Technology},
  publisher = {Scribner},
  year      = {2022},
  isbn      = {978-1982172008}
}

@article{Gereffi2005,
  author    = {Gereffi, Gary and Humphrey, John and Sturgeon, Timothy},
  title     = {The Governance of Global Value Chains},
  journal   = {Review of International Political Economy},
  volume    = {12},
  number    = {1},
  pages     = {78--104},
  year      = {2005},
  doi       = {10.1080/09692290500049805}
}

@article{Farrell2019,
  author    = {Farrell, Henry and Newman, Abraham L.},
  title     = {Weaponized Interdependence: How Global Economic Networks Shape State Coercion},
  journal   = {International Security},
  volume    = {44},
  number    = {1},
  pages     = {42--79},
  year      = {2019},
  doi       = {10.1162/isec_a_00351}
}

@article{Dalkey1963,
  author  = {Dalkey, Norman and Helmer, Olaf},
  title   = {An Experimental Application of the {D}elphi Method to the Use of Experts},
  journal = {Management Science},
  volume  = {9},
  number  = {3},
  pages   = {351--515},
  year    = {1963}
}

@book{Linstone1975,
  editor    = {Linstone, Harold A. and Turoff, Murray},
  title     = {The {Delphi} Method: Techniques and Applications},
  publisher = {Addison-Wesley},
  year      = {1975},
  isbn      = {978-0201042948},
  url       = {https://is.njit.edu/pubs/delphibook/}
}

@techreport{Horton2023,
  author      = {Horton, John J. and Filippas, Apostolos and Manning, Benjamin S.},
  title       = {Large Language Models as Simulated Economic Agents: What Can We Learn from Homo Silicus?},
  institution = {National Bureau of Economic Research},
  type        = {NBER Working Paper},
  number      = {31122},
  address     = {Cambridge, MA},
  year        = {2023},
  note        = {Revised February 2026},
  url         = {http://www.nber.org/papers/w31122}
}

@article{Argyle2023,
  author  = {Argyle, Lisa P. and Busby, Ethan C. and Fulda, Nancy and Gubler, Joshua R. and Rytting, Christopher and Wingate, David},
  title   = {Out of One, Many: Using Language Models to Simulate Human Samples},
  journal = {Political Analysis},
  volume  = {31},
  number  = {3},
  pages   = {337--351},
  year    = {2023},
  doi     = {10.1017/pan.2023.2}
}

@inproceedings{Aher2023,
  author    = {Aher, Gati V. and Arriaga, Rosa I. and Kalai, Adam Tauman},
  title     = {Using Large Language Models to Simulate Multiple Humans and Replicate Human Subject Studies},
  booktitle = {Proceedings of the 40th International Conference on Machine Learning},
  series    = {Proceedings of Machine Learning Research},
  volume    = {202},
  pages     = {337--371},
  year      = {2023},
  publisher = {PMLR}
}

@article{Schoenegger2024,
  author  = {Schoenegger, Philipp and Tuminauskaite, Indre and Park, Peter S. and Valdece Sousa Bastos, Rafael and Tetlock, Philip E.},
  title   = {Wisdom of the Silicon Crowd: {LLM} Ensemble Prediction Capabilities Rival Human Crowd Accuracy},
  journal = {Science Advances},
  volume  = {10},
  number  = {45},
  pages   = {eadp1528},
  year    = {2024},
  doi     = {10.1126/sciadv.adp1528}
}

@techreport{IEA2026kq,
  author      = {{International Energy Agency}},
  title       = {Key Questions on Energy and {AI}},
  institution = {International Energy Agency},
  address     = {Paris},
  year        = {2026},
  url         = {https://www.iea.org/reports/key-questions-on-energy-and-ai}
}

@book{Hirschman1958,
  author    = {Hirschman, Albert O.},
  title     = {The Strategy of Economic Development},
  publisher = {Yale University Press},
  address   = {New Haven},
  year      = {1958}
}

@incollection{HausmannRodrikVelasco2008,
  author    = {Hausmann, Ricardo and Rodrik, Dani and Velasco, Andr\'es},
  title     = {Growth Diagnostics},
  booktitle = {The Washington Consensus Reconsidered: Towards a New Global Governance},
  editor    = {Serra, Narcis and Stiglitz, Joseph E.},
  publisher = {Oxford University Press},
  address   = {Oxford},
  year      = {2008},
  pages     = {324--355},
  doi       = {10.1093/acprof:oso/9780199534081.003.0015}
}

@article{JuhaszLaneRodrik2024,
  author  = {Juh\'asz, R\'eka and Lane, Nathan and Rodrik, Dani},
  title   = {The New Economics of Industrial Policy},
  journal = {Annual Review of Economics},
  volume  = {16},
  pages   = {213--242},
  year    = {2024},
  doi     = {10.1146/annurev-economics-081023-024638}
}

@book{Nardo2008,
  author    = {{OECD} and {European Union} and {EC-JRC}},
  title     = {Handbook on Constructing Composite Indicators: Methodology and User Guide},
  publisher = {OECD Publishing},
  address   = {Paris},
  year      = {2008},
  doi       = {10.1787/9789264043466-en}
}

@misc{EpochAI2026,
  author       = {{Epoch AI}},
  title        = {{AI} Models Documentation},
  year         = {2026},
  howpublished = {Published online at epoch.ai},
  note         = {Frontier models defined as those in the top five by training compute at release},
  url          = {https://epoch.ai/data/ai-models-documentation}
}

@article{BresnahanTrajtenberg1995,
  author  = {Bresnahan, Timothy F. and Trajtenberg, Manuel},
  title   = {General Purpose Technologies: `Engines of Growth'?},
  journal = {Journal of Econometrics},
  volume  = {65},
  number  = {1},
  pages   = {83--108},
  year    = {1995},
  doi     = {10.1016/0304-4076(94)01598-T}
}

@article{Baldwin2022,
  author  = {Baldwin, Richard and Freeman, Rebecca},
  title   = {Risks and Global Supply Chains: What We Know and What We Need to Know},
  journal = {Annual Review of Economics},
  volume  = {14},
  pages   = {153--180},
  year    = {2022},
  doi     = {10.1146/annurev-economics-051420-113737}
}

@article{Aghion2015,
  author  = {Aghion, Philippe and Cai, Jing and Dewatripont, Mathias and Du, Luosha and Harrison, Ann and Legros, Patrick},
  title   = {Industrial Policy and Competition},
  journal = {American Economic Journal: Macroeconomics},
  volume  = {7},
  number  = {4},
  pages   = {1--32},
  year    = {2015},
  doi     = {10.1257/mac.20120103}
}

@techreport{Rodrik2004,
  author      = {{European Commission, Directorate-General for Economic and Financial Affairs}},
  title       = {Industrial Policy for the Twenty-First Century},
  institution = {European Commission},
  type        = {Discussion Paper},
  number      = {157},
  year        = {2022},
  url         = {https://economy-finance.ec.europa.eu/system/files/2022-02/dp157_en_industrial_policy.pdf}
}

@incollection{Agrawal2019,
  author    = {Agrawal, Ajay and Gans, Joshua and Goldfarb, Avi},
  title     = {Economic Policy for Artificial Intelligence},
  booktitle = {Innovation Policy and the Economy},
  volume    = {19},
  publisher = {University of Chicago Press},
  address   = {Chicago},
  pages     = {139--159},
  year      = {2019},
  doi       = {10.1086/699935}
}

@article{Ouyang2025,
  author  = {Ouyang, Shuyin and Zhang, Jie M. and Harman, Mark and Wang, Meng},
  title   = {An Empirical Study of the Non-Determinism of {ChatGPT} in Code Generation},
  journal = {ACM Transactions on Software Engineering and Methodology},
  volume  = {34},
  number  = {2},
  pages   = {1--28},
  year    = {2025},
  doi     = {10.1145/3697010}
}

@techreport{TrendForce2026,
  author      = {{TrendForce}},
  title       = {Revenue Ranking among Top 10 Global Foundries},
  institution = {TrendForce},
  type        = {Quarterly Research Report},
  year        = {2026},
  url         = {https://www.trendforce.com/research/download/RP260608BR3}
}

@misc{ArtificialAnalysis2026,
  author       = {{Artificial Analysis}},
  title        = {Artificial Analysis Economics Index},
  year         = {2026},
  howpublished = {Published online at artificialanalysis.ai},
  url          = {https://artificialanalysis.ai/models/capabilities/economics}
}

@misc{Crunchbase2026,
  author       = {{Crunchbase News}},
  title        = {Billion-Dollar Startup Exits: {M\&A}, {IPO} and {SPAC} Data, {Q2} 2026},
  year         = {2026},
  howpublished = {Published online},
  url          = {https://news.crunchbase.com/public/data-billion-dollar-startup-exits-ma-ipo-spcx-q2-2026/}
}

@techreport{IEA2026elec,
  author      = {{International Energy Agency}},
  title       = {Electricity 2026},
  institution = {International Energy Agency},
  address     = {Paris},
  year        = {2026},
  url         = {https://www.iea.org/reports/electricity-2026}
}

@misc{NVIDIA2026,
  author       = {{NVIDIA Corporation}},
  title        = {{NVIDIA} Announces Financial Results for Second Quarter Fiscal 2027},
  year         = {2026},
  month        = {August},
  howpublished = {Form 8-K, US Securities and Exchange Commission},
  url          = {https://www.sec.gov/Archives/edgar/data/0001045810/000104581026000073/q2fy27pr.htm}
}

@misc{AMD2026,
  author       = {{Advanced Micro Devices}},
  title        = {{AMD} Reports Second Quarter 2026 Financial Results},
  year         = {2026},
  month        = {August},
  howpublished = {Press release, 4 August 2026},
  url          = {https://ir.amd.com/news-events/press-releases}
}

@misc{DeveloperProjections2026,
  author       = {{Fortune}},
  title        = {{OpenAI} Cash Burn Rate and Annual Losses through 2028},
  year         = {2025},
  month        = {November},
  url          = {https://fortune.com/2025/11/12/openai-cash-burn-rate-annual-losses-2028-profitable-2030-financial-documents/}
}

@misc{Meta2026,
  author       = {{CNBC}},
  title        = {Meta Debuts First Major {AI} Model Since \$14 Billion Deal to Bring in Alexandr Wang},
  howpublished = {\url{https://www.cnbc.com/2026/04/08/meta-debuts-first-major-ai-model-since-14-billion-deal-to-bring-in-alexandr-wang.html}, retrieved 3 September 2026},
  year         = {2026}
}

@misc{OpenAI2026round,
  author       = {{OpenAI}},
  title        = {{OpenAI} Raises \$122 Billion to Accelerate the Next Phase of {AI}},
  year         = {2026},
  url          = {https://openai.com/index/accelerating-the-next-phase-ai/}
}

@misc{Anthropic2026,
  author       = {{Anthropic}},
  title        = {Anthropic Raises \$65{B} in Series {H} Funding at \$965{B} Post-Money Valuation},
  year         = {2026},
  url          = {https://www.anthropic.com/news/series-h}
}

@misc{Bloomberg2026,
  author       = {{Bloomberg}},
  title        = {{OpenAI} Valued at \$852 Billion After Completing \$122 Billion Round},
  howpublished = {\url{https://www.bloomberg.com/news/articles/2026-03-31/openai-valued-at-852-billion-after-completing-122-billion-round}, retrieved 3 September 2026},
  year         = {2026}
}

@article{Acemoglu2012,
  author  = {Acemoglu, Daron and Carvalho, Vasco M. and Ozdaglar, Asuman and Tahbaz-Salehi, Alireza},
  title   = {The Network Origins of Aggregate Fluctuations},
  journal = {Econometrica},
  volume  = {80},
  number  = {5},
  pages   = {1977--2016},
  year    = {2012},
  doi     = {10.3982/ECTA9623}
}

@article{Carvalho2019,
  author  = {Carvalho, Vasco M. and Tahbaz-Salehi, Alireza},
  title   = {Production Networks: A Primer},
  journal = {Annual Review of Economics},
  volume  = {11},
  pages   = {635--663},
  year    = {2019},
  doi     = {10.1146/annurev-economics-080218-030212}
}

@article{Baqaee2019,
  author  = {Baqaee, David Rezza and Farhi, Emmanuel},
  title   = {The Macroeconomic Impact of Microeconomic Shocks: Beyond {H}ulten's Theorem},
  journal = {Econometrica},
  volume  = {87},
  number  = {4},
  pages   = {1155--1203},
  year    = {2019},
  doi     = {10.3982/ECTA15202}
}

@article{Graedel2012,
  author  = {Graedel, T. E. and Barr, Rachel and Chandler, Chelsea and Chase, Thomas and Choi, Joanne and Christoffersen, Lee and Friedlander, Elizabeth and Henly, Claire and Jun, Christine and Nassar, Nedal T. and Schechner, Daniel and Warren, Simon and Yang, Man-Yu and Zhu, Charles},
  title   = {Methodology of Metal Criticality Determination},
  journal = {Environmental Science \& Technology},
  volume  = {46},
  number  = {2},
  pages   = {1063--1070},
  year    = {2012},
  doi     = {10.1021/es203534z}
}

@article{Blengini2017,
  author  = {Blengini, Gian Andrea and Nuss, Philip and Dewulf, Jo and Nita, Viorel and Talens Peir\'o, Laura and Vidal-Legaz, Beatriz and Latunussa, Cynthia and Mancini, Lucia and Blagoeva, Darina and Pennington, David and Pellegrini, Mattia and {Van Maercke}, Alexis and Solar, Slavko and Grohol, Milan and Ciupagea, Constantin},
  title   = {{EU} Methodology for Critical Raw Materials Assessment: Policy Needs and Proposed Solutions for Incremental Improvements},
  journal = {Resources Policy},
  volume  = {53},
  pages   = {12--19},
  year    = {2017},
  issn    = {0301-4207},
  doi     = {10.1016/j.resourpol.2017.05.008}
}

@book{NRC2008,
  author    = {{National Research Council}},
  title     = {Minerals, Critical Minerals, and the {U.S.} Economy},
  publisher = {National Academies Press},
  address   = {Washington, DC},
  year      = {2008},
  doi       = {10.17226/12034}
}

\appendix
\section{Appendix}

\subsection{Model sample}

\begin{table}[htbp]
\centering\small
\caption{Frontier sample and capability scores, 31 August 2026}\label{tab:models}
\begin{tabular}{llc}
\toprule
Developer & Model & Score \\
\midrule
Anthropic & Claude Opus 5 & 63 \\
OpenAI & GPT-5.6 Sol & 61 \\
xAI & Grok 4.6 & 61 \\
Moonshot & Kimi K3 & 60 \\
Zhipu & GLM-5.3 & 60 \\
Alibaba & Qwen3.8-Max & 58 \\
Meta & Muse Spark 1.2 & 57 \\
Google & Gemini 3.7 Flash & 56 \\
DeepSeek & DeepSeek V4-Pro & 53 \\
\midrule
\textbf{Average} & & \textbf{58.8} \\
\bottomrule
\end{tabular}
\end{table}

\noindent Scores are author-verified leaderboard values for models released on or before the reference date. All nine lie within ten points of the leader; their average of 58.8 gives $R = 1 - 58.8/100 = 0.412$. A five-point cutoff selects six models and gives $R = 0.395$; thresholds from ten to seventeen points select the same nine, since the tenth-ranked model scores eight points below the ninth; including the two developers immediately below the frontier gives $R = 0.451$. The headline index lies within $[0.754, 0.755]$ under all of them.

\subsection{Monte Carlo}

Robustness intervals apply $\pm 20$ percent uniform perturbation to all parameters except the capped growth ratios, clipped to $[0,1]$, over 10{,}000 draws.

\subsection{Confidence-weighted specification}

Two parameters are rated below medium confidence: $E$ (medium-low) and $SU$ (low). Excluding them and renormalising leaves $RC$ as the sole talent component, $V_T^{adj} = 0.479$, and capital weights of 0.533, 0.200 and 0.267 on $DEP$, $RI$ and $VS$, giving $V_K^{adj} = 0.503$. Under Scenario A weights,
\begin{equation*}
\text{AIIVI}^{adj} = 1 - \left(0.166^{0.35} \times 0.697^{0.05} \times 0.521^{0.10} \times 0.497^{0.20} \times 0.140^{0.30}\right) = 0.763,
\end{equation*}
against the main estimate of 0.754.

\section{Sitography}

Sources below are cited in the text as [W$n$]. All were retrieved on 3 September 2026. Tier 1 denotes an issuer, government or institutional publication; tier 2 a named outlet or research firm reporting a figure originating elsewhere. Sources for the fourteen index parameters are catalogued separately in Supplementary Material 2.

\small
\begin{description}
\item[{[W1]}] Congressional Research Service, \emph{Semiconductors and the CHIPS Act: The Global Context}, report R47558. US CHIPS Act appropriation and European Chips Act mobilisation target. Tier 1.\\ \url{https://www.congress.gov/crs-product/R47558}

\item[{[W2]}] TrendForce, ``AI Inference Chip Architecture''. NVIDIA's licensing of Groq inference technology and engineering team, December 2025. Tier 2.\\ \url{https://www.trendforce.com/insights/ai-inference-chip-architecture}

\item[{[W3]}] Statista, capital expenditure of Meta, Alphabet, Amazon and Microsoft, compiled from company reports. Tier 2, compiling primary filings.\\ \url{https://www.statista.com/chart/35046/capital-expenditure-of-meta-alphabet-amazon-and-microsoft/}

\item[{[W4]}] NVIDIA Corporation, second quarter fiscal 2027 disclosure of supply and capacity commitments and of lease guarantees for the Ohio campus, with terms filed under Item 2.03; as reported by Constellation Research, Jon Peddie Research and Tech Times. Tier 2.\\ \url{https://www.constellationr.com/insights/news/openai-nvidia-sb-energy-finalize-ohio-ai-infrastructure-deal}\\ \url{https://www.jonpeddie.com/news/nvidia-financial-results-for-q226-fy27/}\\ \url{https://www.techtimes.com/articles/324812/20260818/nvidia-filed-105b-ohio-guarantee-not-250b-sec-disclosure-corrects-july-reports.htm}

\item[{[W5]}] Gallup, national survey of 1{,}000 US adults conducted 2--18 March 2026 on local data-centre construction, as reported by ConstructConnect News. Tier 2 reporting a tier 1 survey.\\ \url{https://news.constructconnect.com/gallup-survey-finds-broad-local-opposition-to-data-centers-as-u.s.-construction-surges}

\item[{[W6]}] Monitoring Analytics, independent market monitor for the PJM Interconnection, capacity-price attribution for the 2025/2026 auction, as reported by the Institute for Energy Economics and Financial Analysis. Tier 2 reporting a tier 1 monitor.\\ \url{https://ieefa.org/resources/projected-data-center-growth-spurs-pjm-capacity-prices-factor-10}

\item[{[W7]}] Office of the Governor of New York, Executive Order 62, statewide moratorium on new hyperscale data centres, 14 July 2026. Tier 1.\\ \url{https://www.governor.ny.gov/news/first-statewide-moratorium-new-hyperscale-data-centers-launched-governor-kathy-hochul}

\item[{[W8]}] Utility Dive, Texas interconnection-queue audit; the Governor's statement of 20 August 2026 as reported by the Daily Caller. Tier 2.\\ \url{https://www.utilitydive.com/news/texas-hits-pause-data-center-interconnections/827046/}\\ \url{https://dailycaller.com/2026/08/23/data-center-backlash-brings-thousands-projects-screeching-halt-texas-new-york/}

\item[{[W9]}] Investing.com, hyperscaler free cash flow for the second quarter of 2026. Tier 2.\\ \url{https://www.investing.com/news/stock-market-news/microsoft-only-us-hyperscaler-with-positive-free-cash-flow-as-ai-capex-bites-93CH-4846663}

\item[{[W10]}] Meta Platforms, ``Meta Reports Second Quarter 2026 Results'', 29 July 2026. Tier 1.\\ \url{https://investor.atmeta.com/investor-news/press-release-details/2026/Meta-Reports-Second-Quarter-2026-Results/default.aspx}

\item[{[W11]}] FactSet and Reuters data on hyperscaler debt issuance and secondary bond pricing, as reported at mlq.ai; issuance aggregate from Sage Advisory. Tier 2.\\ \url{https://mlq.ai/news/hyperscaler-bond-issuance-reaches-about-220-billion-as-ai-financing-costs-rise/}\\ \url{https://www.sageadvisory.com/article/hyperscaler-debt-deluge-the-new-driver-of-ig-spread-pressure}

\item[{[W12]}] Financing of Meta's El Paso data-centre campus, as reported by Yahoo Finance; deal structure from Capacity. Tier 2.\\ \url{https://finance.yahoo.com/technology/ai/articles/metas-ai-borrowing-costs-rise-163735686.html}\\ \url{https://capacityglobal.com/news/metas-12bn-el-paso-bond-exposes-ai-debt-repricing/}
\end{description}

\end{document}


\begin{center}
{\large\bfseries Supplementary Material 1}\\[2pt]
Protocol and Elicitation Pilot\\[2pt]
\end{center}

\section*{1. Protocol}

Parameters are produced in three steps. An AI system searches for and extracts the raw inputs to each parameter definition, computes the parameter, and drafts a justification. The authors read each output and either validate it against the original source or reject it, directing revision where required. Validated values enter the manuscript with a confidence rating and a source reference; rejected values return to extraction.

Large language model sessions are not deterministic, so an identical prompt does not in general return an identical output. Reproducibility therefore rests on the data, the source catalogue and the code, all of which are deterministic, rather than on the prompts. The instruments in Supplementary Material 3 document procedure.

\begin{center}
\begin{tikzpicture}[
  font=\small,
  box/.style={draw, rounded corners=2pt, align=center, inner sep=5pt},
  src/.style={box, fill=gray!12, text width=44mm},
  ai/.style={box, fill=blue!8, text width=76mm},
  hum/.style={box, fill=orange!14, text width=76mm},
  ms/.style={box, fill=green!12, text width=44mm},
  rep/.style={box, fill=gray!8, text width=112mm},
  arr/.style={-{Stealth[length=2.6mm]}, thick},
  darr/.style={arr, dashed},
  lbl/.style={font=\scriptsize\itshape, align=center, fill=white, inner sep=1pt}
]
\node[src] (sources) at (0,0) {\textbf{Sources}\\ \scriptsize filings, institutional releases,\\ \scriptsize named outlets};
\node[src, right=10mm of sources] (raw) {\textbf{Raw data}\\ \scriptsize elicitation responses,\\ \scriptsize co-authorship network};
\path (sources.south) -- (raw.south) coordinate[midway] (mid);
\node[ai, below=9mm of mid] (s1) {\textbf{AI-assisted processing}\\ \scriptsize extraction, computation, drafting under structured instruments};
\node[hum, below=11mm of s1] (s2) {\textbf{Author review}\\ \scriptsize validate or reject against the original source; confidence rating};
\node[ms, below=11mm of s2] (msn) {\textbf{Manuscript}\\ \scriptsize validated values};
\node[rep, below=9mm of msn] (rl) {\textbf{Replication layer} --- data, source catalogue, code\\ \scriptsize recomputes every reported figure};
\draw[arr] (sources) -- (s1.north -| sources);
\draw[arr] (raw) -- (s1.north -| raw);
\draw[arr] (s1) -- (s2);
\draw[arr] (s2) -- (msn) node[midway, right=1mm, lbl]{validated};
\draw[arr] (s2.west) .. controls +(-16mm,0) and +(-16mm,0) .. (s1.west) node[midway, left=1mm, lbl]{rejected};
\draw[darr] (raw.east) .. controls +(22mm,-4mm) and +(13mm,10mm) .. (rl.east);
\draw[arr] (rl.north -| msn) -- (msn.south) node[midway, right=1mm, lbl]{recomputes};
\end{tikzpicture}
\end{center}

\section*{2. Confidence ratings}

Each parameter is rated on four criteria: authenticity of the source, accuracy of the extracted figure, correctness of the computation, and validity of the computation given the parameter definition. Ratings run High, Medium-High, Medium, Medium-Low, Low. High requires a primary source, a directly reported figure, and no definitional judgment. Low indicates that no primary disclosure exists and the figure rests on issuer projections. The manuscript reports a specification excluding all parameters rated below Medium.

\section*{3. Elicitation pilot}

\subsection*{3.1 Composition}

Eleven respondents completed an identical instrument: seven human experts in industrial economics and artificial intelligence, and four AI systems from independent developers. Two further AI responses collected in the same wave were excluded before analysis, one from a system built on another sample vendor's model and one from a session in which the study's framing had been given in advance. Results for the full thirteen-response wave are reported alongside the cleaned set.

\subsection*{3.2 Instrument}

Respondents distributed 100 points across the components of each dimension according to non-substitutability, and 100 points across the five dimensions according to the impact of a substantial reduction in each input. Reference weights were disclosed. Weights were normalised within each section before aggregation. The instrument is reproduced in Supplementary Material 3.

\subsection*{3.3 Results}

Applied to the 31 August 2026 parameter set, the cleaned-panel ($n=11$) mean weights give an index of 0.681 and the full wave ($n=13$) gives 0.699. Individual values have mean 0.682 (sd 0.065) and range $[0.592, 0.815]$. Human respondents average 0.658 and AI respondents 0.726, a difference of 0.068 (Welch's $t = 2.24$, $p = 0.056$).

Three departures from the disclosed reference weights are common to most respondents: the carbon-regulation component is weighted lower, the efficiency component higher, and compute criticality lower.

Anonymised responses are available from the authors on request.

\section*{4. Replication package}

\begin{tabular}{ll}
\toprule
\texttt{aiivi\_v30\_replication.py} & Reproduces all reported index values, intervals and variants \\
\texttt{Expert\_Elicitation\_Responses.xlsx} & Raw pilot responses \\
\texttt{neurips2025\_*.py}, \texttt{*.csv} & Co-authorship network extraction and output \\
Supplementary Material 2 & Parameter sources and derivations \\
Supplementary Material 3 & Instruments and code index \\
\bottomrule
\end{tabular}

\medskip
\noindent Scripts run on Python 3.11 with \texttt{numpy}, \texttt{pandas}, \texttt{scipy} and \texttt{openpyxl}.


\begin{center}
{\large\bfseries Supplementary Material 2}\\[2pt]
Parameter Sources and Derivations\\[2pt]
\end{center}

\section*{1. Conventions}

\textbf{Reporting period.} Each firm-level parameter uses the most recent period ended on or before 31 August 2026, taking the disclosure as published. Where a disclosure appeared after the reference date but covers a period ending before it, the period governs.

\textbf{Aggregation.} Parameters computed across firms are formed from industry totals, not from averages of firm-level ratios, so that firms enter in proportion to economic weight.

\textbf{Sources.} Official issuer publications, institutional statistical releases, and named outlets with live article URLs. Source tier enters the confidence rating: tier 1 denotes issuer or institutional publication, tier 2 a named outlet reporting a figure originating elsewhere.

\textbf{Scope.} This catalogue covers the fourteen index parameters. Sources for contextual claims in the manuscript's discussion are footnoted at the point of use, with publisher, URL and retrieval date.

\section*{2. Derivations}

\begin{longtable}{@{}lp{12.4cm}@{}}
\toprule
\textbf{Parameter} & \textbf{Value, computation and sources} \\
\midrule
\endhead
$SC_{comp}$ & $0.732$. $\sqrt{\sum_k s_k^2}$ over top-ten foundry revenue shares for the second quarter of 2026, the most recent quarter ended on or before the reference date: TSMC 72.5, Samsung 5.9, SMIC 5.4, UMC 3.9, GlobalFoundries 3.2, HuaHong 2.3, and Tower, VIS, Nexchip and PSMC at 0.8 each (S1, S1b). The preceding quarter gives 0.730. \\
\addlinespace
$GEO_{comp}$ & $0.920$. Share of manufacturing capacity for the most advanced nodes located in Taiwan (S2). Annual cadence. \\
\addlinespace
$PR_{comp}$ & $0.793$. $\sqrt{\sum_k s_k^2}$ over disclosed data-centre segment revenue of the four suppliers reporting a comparable segment, each for its most recent period ended on or before the reference date: NVIDIA \$89.0B, quarter ended 26 July 2026 (S3); Broadcom \$16.7B AI semiconductor revenue, quarter ended 2 August 2026 (S4); AMD \$6.7B, quarter ended June 2026 (S5); Marvell \$2.17B, quarter ended 1 August 2026 (S6). Shares 0.777/0.146/0.059/0.019. The four periods end within five weeks. Segment definitions bundle compute with networking at all four firms, so the measure is symmetric across them. Revenue proxies deployed compute, which no vendor discloses. Variant, merchant vendors alone: 0.931. \\
\addlinespace
$GR_{energy}$ & $1.000$. Electricity demand at AI-focused data centres grew 50 percent in 2025 against 3 percent for global electricity demand (S7). Demand growth exceeds supply growth, so the parameter takes its normalised maximum. \\
\addlinespace
$GR_{co2}$ & $1.000$. AI-related emissions growth exceeds global emissions growth (S7, S8). Normalised maximum as above. \\
\addlinespace
$EB_{energy}$ & $0.533$. Like-for-like training basis: $1 - 1.75/3.75$, where 3.75 is the Blackwell generation's efficiency gain over its predecessor and 1.75 the Rubin generation's, computed as 3.5$\times$ throughput (S10) at approximately twice the per-GPU power draw (S9). Vendor rack-scale per-watt claims rest on reduced-precision arithmetic (S11) and are excluded from the like-for-like series. Variants: 0.333 and 0.627 on alternative throughput bases. \\
\addlinespace
$E_{talent}$ & $0.458$. Concentration of elite researchers across institutions (S12). Annual cadence. \\
\addlinespace
$RC_{talent}$ & $0.479$. Sum of normalised betweenness centrality of the five highest-ranked corporate institutions in the co-authorship network of the leading machine-learning venue: 0.228, 0.127, 0.065, 0.034, 0.025 (S13). Extraction code and network output are in the replication package. Annual cadence. \\
\addlinespace
$DEP_{capital}$ & $0.590$. $\sqrt{\sum_k s_k^2}$ over unconditional strategic equity from any source. NVIDIA \$40B (\$30B in OpenAI, S14--S15, reported as not tied to deployment milestones, S18; \$10B in Anthropic, S17); Amazon \$28B (\$15B unconditional of a \$50B commitment, S16; \$8B and \$5B in Anthropic, S17); Microsoft \$18.75B; Google \$2B. Shares 0.451/0.315/0.211/0.023. SoftBank's \$30B is excluded as a financial rather than strategic investor. Variant, cloud providers only: 0.693. \\
\addlinespace
$SU_{capital}$ & $0.104$. Industry totals over the developers with published burn-to-revenue projections (S19). OpenAI 57 percent of revenue on \$25B annualised; Anthropic approximately one third on \$47B annualised. Aggregate burn \$29.9B against four years' aggregate revenue of \$288B. No frontier developer publishes audited accounts, so the parameter rests on issuer projections; confidence Low. Variant, firm averages: 0.113. \\
\addlinespace
$RI_{bubble}$ & $0.316$. Industry totals over the developers whose valuations rest on disclosed rounds: aggregate capitalisation \$1{,}817B (OpenAI \$852B, S14; Anthropic \$965B, S17) on aggregate annualised revenue \$72B, giving 25.2$\times$ against the 80$\times$ benchmark. Cap inactive. Developers whose market data cannot be sourced to filings or issuer publications are excluded. Variant, firm averages: 0.341. \\
\addlinespace
$VS_{capital}$ & $0.465$. Foundation model developer raises of \$237B against \$510B in global venture funding in the first half of 2026 (S20). \\
\addlinespace
$R_{data}$ & $0.412$. $1 - 58.8/100$, the mean capability score of the nine-model frontier set (S21). Frontier defined as models within ten index points of the leader; the model-tracking convention is a top-$N$ cut on training compute at release (S23), which is not disclosed for these models, so the cut is applied on measured capability. Variants: 0.395 at a five-point cutoff, 0.451 including the two developers immediately below the frontier. \\
\addlinespace
$C_{data}$ & $0.049$. Training-cost economics relative to developer revenue (S22). Annual cadence. \\
\bottomrule
\end{longtable}

\section*{3. Sources}

\noindent\textbf{S1} TrendForce, \emph{Revenue Ranking among Top 10 Global Foundries}, quarterly research report, tier 1. \url{https://www.trendforce.com/research/download/RP260608BR3}

\noindent\textbf{S1b} TrendForce 2Q26 foundry ranking table as reported by Evertiq, tier 2. \url{https://evertiq.com/design/2026-09-10-smic-narrows-gap-with-samsung-as-global-foundry-revenue-hits-record-5349-billion}

\noindent\textbf{S2} United States International Trade Commission, \emph{Silicon Island: Taiwan's Semiconductor Industry}, Executive Briefing on Trade, tier 1. \url{https://www.usitc.gov/publications/332/executive_briefings/ebot_silicon_island_taiwan_semiconductor.pdf}

\noindent\textbf{S3} NVIDIA Corporation, results for the second quarter of fiscal 2027, Form 8-K, 26 August 2026, tier 1. \url{https://www.sec.gov/Archives/edgar/data/0001045810/000104581026000073/q2fy27pr.htm}

\noindent\textbf{S4} Broadcom Inc., results for the third quarter of fiscal 2026, tier 1. \url{https://investors.broadcom.com/news-releases}

\noindent\textbf{S5} Advanced Micro Devices, second quarter 2026 results, 4 August 2026, tier 1. \url{https://ir.amd.com/news-events/press-releases}

\noindent\textbf{S6} Marvell Technology, results for the second quarter of fiscal 2027, 27 August 2026, tier 1. \url{https://investor.marvell.com/news-releases}

\noindent\textbf{S7} International Energy Agency, \emph{Key Questions on Energy and AI}, April 2026, tier 1. \url{https://www.iea.org/reports/key-questions-on-energy-and-ai}

\noindent\textbf{S8} International Energy Agency, \emph{Electricity 2026}, February 2026, tier 1. \url{https://www.iea.org/reports/electricity-2026}

\noindent\textbf{S9} CNBC, 25 February 2026, reporting NVIDIA's statement on Vera Rubin power draw, tier 2. \url{https://www.cnbc.com/2026/02/25/first-look-at-nvidias-ai-system-vera-rubin-and-how-it-beats-blackwell.html}

\noindent\textbf{S10} VentureBeat, 13 January 2026, vendor throughput claims as reported, tier 2. \url{https://venturebeat.com/infrastructure/nvidias-vera-rubin-is-months-away-blackwell-is-getting-faster-right-now}

\noindent\textbf{S11} NVIDIA Technical Blog, Rubin platform architecture, tier 1. \url{https://developer.nvidia.com/blog/inside-the-nvidia-rubin-platform-six-new-chips-one-ai-supercomputer/}

\noindent\textbf{S12} Stanford Institute for Human-Centered AI, \emph{AI Index Report} 2026, tier 1. \url{https://arxiv.org/abs/2606.15708}

\noindent\textbf{S13} OpenReview, NeurIPS 2025 accepted-paper metadata; author extraction and network construction in the replication package, tier 1.

\noindent\textbf{S14} OpenAI, ``Accelerating the Next Phase of AI'', 31 March 2026, tier 1. \url{https://openai.com/index/accelerating-the-next-phase-ai/}

\noindent\textbf{S15} CNBC, 31 March 2026, round composition, tier 2. \url{https://www.cnbc.com/2026/03/31/openai-funding-round-ipo.html}

\noindent\textbf{S16} Bloomberg, 31 March 2026, tranche structure of the Amazon commitment, tier 2. \url{https://www.bloomberg.com/news/articles/2026-03-31/openai-valued-at-852-billion-after-completing-122-billion-round}

\noindent\textbf{S17} Anthropic, Series H announcement, 28 May 2026, tier 1. \url{https://www.anthropic.com/news/series-h}

\noindent\textbf{S18} CNBC, 4 March 2026, reporting that NVIDIA's \$30B investment is not tied to deployment milestones, tier 2. \url{https://www.cnbc.com/2026/03/04/nvidia-huang-openai-investment.html}

\noindent\textbf{S19} Fortune, OpenAI cash burn and annual losses through 2028, 12 November 2025, tier 2. \url{https://fortune.com/2025/11/12/openai-cash-burn-rate-annual-losses-2028-profitable-2030-financial-documents/}

\noindent\textbf{S20} Crunchbase News, billion-dollar startup exits and venture funding data, Q2 2026, tier 2. \url{https://news.crunchbase.com/public/data-billion-dollar-startup-exits-ma-ipo-spcx-q2-2026/}

\noindent\textbf{S21} Artificial Analysis, Economics Index; scores author-verified for models released on or before the reference date, tier 1. \url{https://artificialanalysis.ai/models/capabilities/economics}

\noindent\textbf{S22} Epoch AI, training-cost estimates for notable models, tier 1. \url{https://epoch.ai/data/ai-models-documentation}

\noindent\textbf{S23} Epoch AI, AI Models Documentation (frontier models defined as the top five by training compute at release), tier 1. \url{https://epoch.ai/data/ai-models-documentation}


\begin{center}
{\large\bfseries Supplementary Material 3}\\[2pt]
Instruments and Code\\[2pt]
\small Manuscript v4.0, reference date 31 August 2026
\end{center}

\section*{1. Disclosure}

Extraction and elicitation were carried out with Claude Opus 5 (Anthropic) through a chat interface, in which sampling settings are not user-controlled. The elicitor is selected by a stated rule: the highest-scoring model accessible to the authors on the economics-specific capability index published by the leaderboard provider, where it scores 63 of 100. The eliciting model is a product of a firm in the sample; the conflict is disclosed here and in the manuscript.

For a citable re-estimation, the instruments below should be run through an API with a pinned model string, a documented temperature and a recorded date.

\section*{2. Extraction instrument}

Run once per parameter.

\begin{promptbox}
You are assisting a Stage-1 information extraction under a human-in-the-loop protocol. Every output you produce will be verified by a human against the original source; do not assert anything you cannot attach a URL to.

Parameter: [NAME]. Definition: [FORMULA]. Reference date: [DATE] --- use only information observable on or before this date.

Task: (1) run at least three independent web searches; (2) extract the raw inputs to the formula, quoting the exact figure and naming publisher and publication date for each; (3) prefer primary sources (official releases, major financial press, institutional statistics); mark anything else Tier 2; (4) compute the parameter, showing arithmetic; (5) state a confidence rating (High / Medium-High / Medium / Medium-Low / Low) with one sentence of justification; (6) list every assumption or definitional choice a human verifier must confirm. If sources conflict, report the range; do not silently pick one.
\end{promptbox}

\section*{3. Elicitation instrument}

Run once per elicitation. The same instrument was administered to the pilot panel.

\begin{promptbox}
You are acting as an expert panellist in a structured weight elicitation for the AI Industrial Vulnerability Index (AIIVI). AIIVI measures systemic vulnerability in the foundation model industry using an O-Ring production function: inputs are complements, and a failure in any critical input can halt production regardless of the others.

Reference date for all judgments: [DATE]. Judge substitutability on a one-to-three-year horizon.

Assign two kinds of weights. (A) Within each of the five dimensions (Compute, Data, Talent, Capital, Energy), distribute 100 points across the components listed below according to non-substitutability: a component is weighted higher the harder it is to replace if it becomes unavailable. (B) Across the five dimensions, distribute 100 points according to the impact on foundation model production if that input were reduced by 50 percent. Weights price the impact of an input reduction, not its probability; likelihood is carried by the index parameters.

Components. Compute: SC (leading-edge fabrication concentration), GEO (geographic concentration of leading-edge capacity), PR (data-centre silicon concentration). Energy: GR\_energy (grid capacity against AI demand growth), GR\_co2 (carbon and regulatory constraint), EB (deceleration of energy-efficiency improvement). Talent: E (elite researcher concentration), RC (corporate brokerage of research knowledge flows). Capital: DEP (strategic-investor dependency), SU (operating sustainability), RI (valuation risk), VS (share of global venture capital). Data: R (model quality headroom), C (training-cost economics).

Reference values, disclosed for transparency: Compute SC 25, GEO 50, PR 25; Energy GR\_energy 50, GR\_co2 40, EB 10; Talent E 59, RC 41; Capital DEP 40, SU 35, RI 15, VS 10; Data R 80, C 20; Criticality Compute 40, Energy 30, Capital 15, Talent 10, Data 5. You may adjust from or reject any of these; justify every departure with evidence dated on or before the reference date.

Output: for each section, integer weights that are multiples of 5 and sum to exactly 100, each with a one-to-three-sentence economic justification citing specific dated developments. Then the five criticality weights on the same basis. Finally, the three developments that most changed your judgment, with sources. Do not consult or attempt to reproduce any published AIIVI results.
\end{promptbox}

\section*{4. Code}

\begin{tabular}{ll}
\toprule
\texttt{aiivi\_v30\_replication.py} & Index, sub-indices, Monte Carlo intervals, marginal effects, \\
 & robustness variants, pilot-panel statistics \\
\texttt{neurips2025\_extract.py} & OpenReview metadata extraction \\
\texttt{neurips2025\_network.py} & Co-authorship network construction \\
\texttt{neurips2025\_cbc.py} & Betweenness centrality and corporate aggregation \\
\bottomrule
\end{tabular}

\medskip
\noindent Parameter values are declared at the head of the replication script and documented in Supplementary Material 2. Python 3.11 with \texttt{numpy}, \texttt{pandas}, \texttt{scipy}, \texttt{openpyxl}.